\documentclass[aps,prb,twocolumn,showpacs]{revtex4}
\usepackage[dvips]{graphicx}
\usepackage{amsmath}
\usepackage{amssymb}
\usepackage{amsthm}
\usepackage{bm}

\begin{document}

\title{An Improved Nonperturbative Method for Studying Two-dimensional Vortex Liquids}

\author{J. Yeo}
\author{H. Park}
\author{S. Yi}

\affiliation{Department of Physics, Konkuk University, Seoul 143-701, Korea}

\date{\today}

\begin{abstract}
We show how a systematic improvement can be made on the nonperturbative 
parquet approximation method which was previously
used to study the effect of thermal fluctuations in 
vortex liquids in high-temperature superconductors. 
This is achieved by including an infinite subset of Feynman diagrams
contributing to the renormalized four-point vertex function of the Ginzburg-Landau
model, which was omitted in the original approximation. 
We find that the growing crystalline order in the vortex liquid
is more pronounced in the improved approximation. In particular,
the second and third peaks in the liquid structure factor, which appeared as one peak
in the original approximation, are now resolved. 
\end{abstract}



\maketitle

\section{Introduction}

Thermal fluctuations play a more important role in
high-temperature superconductors than in conventional
materials because of strong anisotropy,
high temperature, and short coherence length \cite{blatter}. 
Therefore, a high-temperature superconductor in a
magnetic field is believed to be in a vortex liquid phase resulting from
the melting of the Abrikosov vortex lattice predicted by the mean
field theory \cite{abrikosov}. 
For bulk materials, experiments \cite{exp1,exp2,exp3}
detect well below the upper critical field $H_{c2}$
sharp drops in resistivity and steps in the magnetization
and the specific heat, which are interpreted when
the strength of disorder is weak as being due to a vortex liquid 
undergoing a first-order phase transition into presumably 
the Abrikosov lattice as the temperature is lowered.
However, the situation is more complicated because the first-order 
transition disappears at both high and low magnetic 
fields \cite{exp1,exp4,exp5,exp6}.

For two-dimensional superconducting films,
the question whether the
vortex liquid undergoes a phase transition at all into
the low-temperature ordered state is still controversial.
The perturbation expansions around the high-temperature liquid state
using the Ginzburg-Landau (GL)
model within the lowest Landau level (LLL) approximation have been carried out 
for evidence of the transition to the ordered phase 
at low temperatures \cite{rt,hmm}. 
A weak first-order transition is detected in two-dimensional vortex liquids 
from numerical simulations \cite{sim}.
However, numerical simulation results depend strongly on the boundary conditions
as the Monte-Carlo simulation performed
on a spherical geometry \cite{om} shows no sign of a
finite-temperature transition.

As a nonperturbative analytic (i.\ e.\ nonsimulational) approach,
the parquet approximation has been successfully applied
to a two-dimensional vortex system \cite{ym} and also
to vortices in a layered superconductor \cite{ym2,yeo}. 
It is free from any finite-size effect perpendicular to
the field direction and sophisticated enough to capture 
the growing crystalline order developing in the vortex
liquid as the temperature is lowered. The parquet approximation
deals with the renormalized four-point function of
the vortex system which is obtained by 
summing an infinite subset of Feynman diagrams,
the so-called parquet diagrams. 
Although the parquet diagrams seem to form a minimal set
of Feynman diagrams which capture the growing crystalline order
properly, there is no a priori reason to neglect the non-parquet
diagrams. It is because there is no apparent small parameter 
associated with the non-parquet diagrams. In this paper, we present
a first attempt to go beyond the parquet approximation. We show 
how one can include systematically the non-parquet diagrams into the existing 
nonperturbative calculations. We do this by devising a way to take
into account yet another infinite subset of Feynman diagrams which
are omitted in the previous calculations. This procedure is similar to what one
does in the integral equation approach to ordinary liquids, \cite{hansen}
where integral equations such as the hypernetted chain equation
are improved by adding an appropriate set of diagrams.
We shall see that 
the growing crystalline order developing in the vortex liquid as the temperature 
decreases is more pronounced 
when the non-parquet diagrams are included, although
no finite temperature phase transition is detected as in the previous
studies \cite{ym}.

In the next section, we briefly set up the parquet diagram decomposition method
for the two-dimensional vortex liquid systems. In Sec.~\ref{diagrams},
we show how the non-parquet contributions can be included in the formulation.
The improved integral equations which include these new diagrams are solved
numerically in Sec.~\ref{results}. We conclude with discussion in Sec.~\ref{discussion}.

\section{Parquet Graph Resummation Method}

The parquet graph resummation method for vortex liquids \cite{ym} is based on the
LLL approximation of the GL model for a 
superconductor in a perpendicular magnetic field.
For a superconducting film, the GL free energy 
with the order parameter denoted by $\Psi$,
is given by
\begin{eqnarray}
F[\Psi] &=& \int d^2 {\bf r} \;
\Big[ \alpha |\Psi({\bf r})|^2
+\frac{\beta}{2}|\Psi ({\bf r})|^4  \nonumber \\
&&+\frac{1}{2m}\left|\left(-i\hbar\nabla-\frac{e^*}{c}{\bf A}\right)\Psi\right|^2
\Big] , 
\end{eqnarray}
where $\alpha$, $\beta$, and $m$ are phenomenological parameters and $e^*=2e$.
We take ${\bf B}=\nabla\times{\bf A}$ as constant and uniform, and
use the LLL approximation, which is believed 
to be valid over a large portion of the vortex-liquid region \cite{lll}.
In the symmetric gauge, where ${\bf A}=\frac{B}{2}(-y,x,0)$,
the LLL wavefunction is given by
$\Psi^{\rm LLL} ({\bf r})=\exp(-\mu^2 |z|^2/4)\phi(z)$ where
$\mu^2 = e^* B /\hbar c$ and 
$\phi(z)$ is an arbitrary analytic function of $z=x+iy$.
In the LLL approximation, the free energy becomes
\begin{eqnarray}
F[\phi,\phi^*] &=& 
\int dz^* dz \; \left[ \alpha_H e^{-\mu^2 |z|^2/2}|\phi(z)|^2  \right. \nonumber \\
&&\left. + \frac{\beta}{2} e^{-\mu^2 |z|^2} |\phi(z)|^4  \right]
\end{eqnarray}
where $\alpha_H \equiv \alpha + e^* B \hbar /2c m $ changes sign
crossing the upper critical field line $H_{c2} (T)$. 
The effect of thermal fluctuations in the two-dimensional vortex liquid
systems is determined by the partition function
$
Z=\int  \mathcal {D}\phi {\cal D}\phi^* \exp (-F[\phi,\phi^*]) . 
$
In this section, we set up the parquet diagram resummation method 
and calculate various correlation functions with respect to the
partition function.
One can develop the standard perturbation theory 
from the given partition function.
The bare propagators are given by
\begin{equation}
\langle\phi^* (z^{\prime *}) \phi (z)\rangle_0 
= \frac 1 {\alpha_H} \left(\frac{\mu^2}{2\pi}\right) e^{\mu^2 z^{\prime *}z/2}. 
\label{propagator}
\end{equation}
Since the magnetic length $\mu^{-1}$ is the only length scale 
perpendicular to the field direction which appears 
in the propagator \cite{ym},
the fully renormalized propagator can also be written in the same way
as Eq.~(\ref{propagator})
with the renormalized $\alpha_R$ replacing the bare $\alpha_H$.
The renormalized $\alpha_R$ is determined self-consistently in the
parquet approximation as will be seen later.

The main quantity one calculates in the parquet approximation
is the renormalized connected  four-point function
$
\langle\phi^*(z_1^*)\phi^*(z_2^*)\phi(z_3)\phi(z_4)\rangle_c .
$ 
An important feature of the LLL approximation is that this renormalized 
quantity can be completely described by a single vertex function $
\Gamma ({\bf k})$ \cite{ym}, where the momentum ${\bf k}$ corresponds 
to the two-dimensional space perpendicular to the magnetic field. 
In general this quantity can be written as
\begin{eqnarray}
&&\langle\phi^* (z^*_1)\phi^* (z^*_2)
\phi (z_3)\phi (z_4)\rangle _c = -
\frac{2\beta}{\alpha_R^4}\left(\frac{\mu^2}{2\pi}\right)^2   \label{gen:four} \\
&&\times\exp\left(\frac{\mu^2}{2}(z^*_1 z_3+z^*_2 z_4)\right) 
\int \frac{dk^*dk}{(2\pi)^2}\;
\Gamma ({\bf k}) \nonumber \\
&&\times\exp \left(-\frac{|k|^2}{2\mu^2}-\frac{i}{2}
\left\{k^*(z_3-z_4)+k(z^*_1-z^*_2)\right\}\right)  ,
\nonumber 
\end{eqnarray}
Note that to the lowest
order, $\Gamma ({\bf k})=
\Gamma_B({\bf k})=\exp(-|k|^2/2\mu^2)$ is the bare vertex.

\begin{figure}
\includegraphics[width=0.425\textwidth]{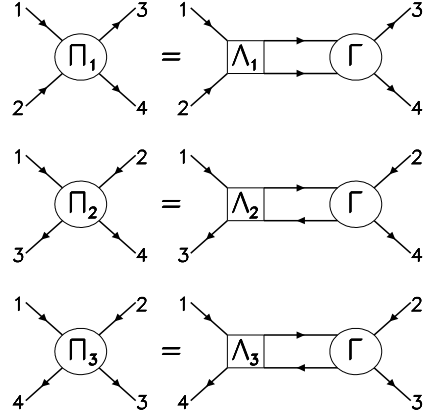}%
\caption{A graphical representation of the Bethe-Salpeter equations 
for the reducible parts $\Pi_i$. (See Eq.~(\ref{o3}).) }
\label{figparq}
\end{figure}

The parquet approximation has been widely used \cite{parquet} in many branches
of many-body physics. For the vortex liquids,
we make a resummation over all parquet diagrams by first noting that
the contributions to $\Gamma$ can be decomposed into the totally 
irreducible part denoted by $R$
and the reducible part. The reducible part in turn can be written
as the sum of three parts $\Pi_i$, ($i=1,2,3)$ 
representing the contributions 
from three different channels as shown in Fig.~\ref{figparq}. 
(A detailed discussion of the diagrammatic decomposition can
be found in Ref.~\onlinecite{ym}.) 
We have 
\begin{equation}
\Gamma({\bf k})=R ({\bf k})+\sum_{i=1}^3 \Pi_i ({\bf k}).
\label{o1}
\end{equation} 
Each reducible vertex $\Pi_i$ is composed of an irreducible vertex
$\Lambda_i$ where  
\begin{equation}
\Lambda_i ({\bf k})=R({\bf k}) +
\sum_{j\neq i}\Pi_j({\bf k}) \label{o2}
\end{equation} 
and the renormalized $\Gamma$ via the following Bethe-Salpeter equations 
(see Fig.~\ref{figparq}):
\begin{subequations}
\label{o3}
\begin{eqnarray}
&&\Pi_1 ({\bf k})=
-x \left[\Lambda_1 \circ \Gamma \right] ({\bf k}), \label{o31} \\
&&\Pi_2 ({\bf k})=
-2x\Lambda_2 ({\bf k})\Gamma ({\bf k}), \label{o32} \\
&&\Pi_3 ({\bf k})=
-2x \left[\Lambda_3 \ast\Gamma\right]({\bf k}), \label{o33} 
\end{eqnarray}
\end{subequations}
where the operation $\circ$
between two arbitrary functions
$f({\bf k})$ and $g({\bf k})$ is defined by
\begin{eqnarray*}
(f\circ g)({\bf k})&=&\frac{2\pi}{\mu^2}\int\frac{
d^2 {\bf k}^\prime}{(2\pi)^2}\; f({\bf k}-{\bf k}^\prime)
g({\bf k}^\prime) \\
&&\times\cos((k_x k^\prime_y -k_y k^\prime_x)/\mu^2)
\end{eqnarray*}
and $f\ast g$ is just the convolution without the cosine factor.
The parquet approximation corresponds
to keeping only the bare vertex function $\Gamma_B ({\bf k})$ 
in the totally irreducible vertex $R({\bf k})$, and 
neglecting all the higher order non-parquet diagrams. The lowest order of the diagrams
neglected in this approximation is O($\beta^4$).  The main point of this paper,
which will be discussed in the next section, is 
to find a way to incorporate systematically the diagrams 
neglected in the parquet approximation.
In Eq.~(\ref{o3}), we have used the 
dimensionless parameter $x=\mu^2\beta/2\pi\alpha^2_R$,
which originated from the two renormalized propagator lines in $\Pi_i$.
This parameter $x$ is determined 
self-consistently from the following {\em exact} Dyson equation, which relates 
the renormalized propagator and the renormalized quartic vertex:
\begin{eqnarray}
\alpha_T&=&\frac{1}{\sqrt{x}}\bigg[ 1-2x+2x^2 \left( \frac{2\pi}{\mu^2}\right)
\int\frac{d^2{\bf k}}{(2\pi)^2}\nonumber\\
&&\quad \quad       \times e^{-k^2/2\mu^2}
\Gamma({\bf k})\bigg],
\label{dyson}
\end{eqnarray}
where $\alpha_T\equiv\alpha_H\sqrt{2\pi/\beta\mu^2}$ is the dimensionless
temperature. We can show that $\alpha_{T}$ is proportional to 
$-(1-t-h)/(th)^{1/2}$, where $t=T/T_{c0}$ and $h=H/H_{c2}(0)$
with $T_{c0}$ and $H_{c2}(0)$ being the critical temperature at 
zero field and the upper critical field at zero temperature, 
respectively.  Note that Eqs.~(\ref{o1}), (\ref{o2}), (\ref{o3}) and (\ref{dyson})
form a closed set of equations for $\Gamma({\bf k})$ for given $\alpha_T$ and 
$R({\bf k})$. Note also that this set of equations are \textit{exact} relations
for the vertex function $\Gamma({\bf k})$.
Recall that in the parquet approximation $R({\bf k})$ is approximated 
as the bare vertex $\Gamma_B({\bf k})$ .

Using the solutions to the above equations we can calculate several
interesting physical quantities. Among them, we focus 
on the structure factor, which
is a measure of the correlation between vortices in a vortex
liquid. It is calculated from 
\begin{equation}
\chi ({\bf r}-{\bf r}^\prime)=\langle
|\Psi ({\bf r})|^2 |\Psi ({\bf r}^\prime)|^2\rangle
-\langle|\Psi ({\bf r})|^2 \rangle\langle
\Psi ({\bf r}^\prime)|^2\rangle .
\end{equation}
The structure factor $\Delta ({\bf k})$
used in this paper is then defined as
\begin{equation}
\Delta ({\bf k}) \equiv 
\left(\frac{2\pi \alpha^2_R}{\mu^2}\right) e^{{\bf k}^2/2\mu^2}
\int d^2 {\bf R} e^{i {\bf k}\cdot{\bf R}} \chi ({\bf R}). 
\label{structure}
\end{equation}
By joining two external legs of the four-point
correlation functions in (\ref{gen:four}), we obtain a simple relation,
\begin{equation}
\Delta({\bf k})=1- 2x \Gamma ({\bf k}). \label{dfv}
\end{equation}
The above coupled integral equations for $\Gamma({\bf k})$ can be solved numerically. 
In Ref.~\onlinecite{ym}, the parquet equations were solved for 
two-dimensional vortex liquids 
with and without quenched impurities. The parquet equations
can also be solved for the vortices in layered superconductors \cite{ym2}.

\section{Non-parquet Contributions}
\label{diagrams}

\begin{figure}
\includegraphics[width=0.45\textwidth]{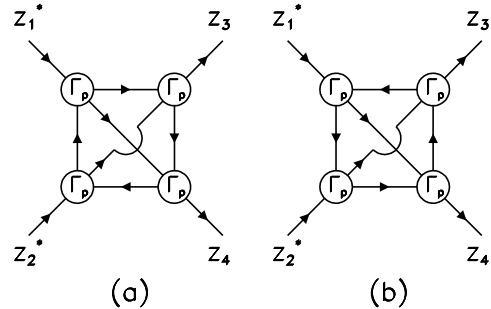}%
\caption{The leading-order non-parquet diagrams. Note that a general
vertex $\Gamma_p({\bf k})$ is used on each of four vertices. There are 
two more diagrams obtained by exchanging $z_3$ and $z_4$.}
\label{np}
\end{figure}

In this section we discuss how the contributions from the non-parquet 
diagrams can be included into the correlation functions of the 
two-dimensional vortex liquid.
Recall that, in the parquet approximation, all the diagrams contributing to 
the totally irreducible vertex $R({\bf k})$ are neglected except
the bare vertex diagram. The lowest order of the 
non-parquet diagrams is $O(\beta^4)$ as shown in Fig.~\ref{np}. 
A straightforward way to proceed would be to calculate the 
non-parquet diagrams term by term starting from the fourth-order diagram.
This procedure will generate a perturbation series for $R({\bf k})$. 
To extract nonperturbative information for $R({\bf k})$ and
for the structure factor, one must then apply a resummation
method such as the Pad\'e approximation to the perturbation series.
Such numerical resummation procedures, however, 
are known \cite{hmm} to be less effective in capturing 
the growing crystalline order
in the vortex liquid than the integral equation
approach like the parquet approximation. 

In this paper, we take a different route by
incorporating an infinite subset of the non-parquet diagrams 
systematically into the integral equations 
discussed in the previous section. 
We first evaluate the diagrams represented by
the lowest order non-parquet diagram shown in 
Fig.~\ref{np}. Note that, in place of the bare vertices, we use a general
vertex $\Gamma_p ({\bf k})$ to be specified later. For any vertex
$\Gamma_p ({\bf k})$, these diagrams certainly 
represent a subset of Feynman diagrams that is not considered in the 
parquet approximation \textit{i.\ e.\ }that cannot be decomposed as
in Fig.~\ref{figparq}. 
We find that there are two distinct diagrams as shown in Fig.~\ref{np},
which turn out to give the same contribution denoted here by $J({\bf k})$.
By explicitly evaluating these diagrams using a general vertex $\Gamma_p ({\bf k})$,
we obtain
\begin{eqnarray}
&&J({\bf k})=-16 x^3 \left( \frac{2\pi}{\mu^2}\right)^2
\int\frac{d^2{\bf p} }{(2\pi)^2}\int\frac{d^2{\bf q}}{(2\pi)^2} \label{J}\\
&&\times \Gamma_p({\bf k}-{\bf q})\Gamma_p({\bf k}-{\bf p})\Gamma_p({\bf p})\Gamma_p({\bf q})
\cos\left(\frac{p_x q_y -p_y q_x}{\mu^2} \right) \nonumber
\end{eqnarray}
There are also contributions from the diagrams obtained by exchanging
$z_3$ and $z_4$ in Fig.~\ref{np}, which are related to $J({\bf k})$
via the hat-operation defined by
\begin{equation}
\widehat{J}({\bf k})=\frac{2\pi}{\mu^2}\int\frac{d^2{\bf p}}{(2\pi)^2}\; J({\bf p})
\cos\left(\frac{k_x p_y -k_y p_x}{\mu^2} \right).
\end{equation}

\begin{figure}
\includegraphics[width=0.25\textwidth]{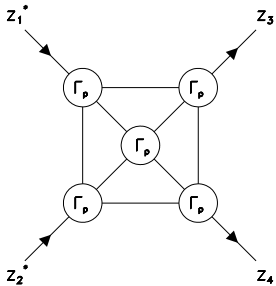}%
\caption{A higher-order non-parquet diagram which can be included 
using the present method}
\label{npn}
\end{figure}

To include the contributions from the new set of diagrams, we take in 
Eqs.~(\ref{o1}) and (\ref{o2})
\begin{equation}
R({\bf k})=\Gamma_B({\bf k}) +J({\bf k})+\widehat{J}({\bf k}).
\label{newR}
\end{equation}
Now we must specify what $\Gamma_p({\bf k})$ is in Eq.~(\ref{J}) to close
the self-consistent equations. In this paper, we take
$\Gamma_p({\bf k})$ as the sum of all the parquet diagrams, that is
the solution of the parquet equations  
(\ref{o1}), (\ref{o2}) and (\ref{o3}) when $R({\bf k})$ is 
just $\Gamma_B({\bf k})$ for given parameter $x$. 
Then the non-parquet diagrams in Fig.~\ref{np}
represent an infinite subset of Feynman diagrams for which the skeleton 
diagram in Fig.~\ref{np} contain
all the parquet diagrams on each of the four vertices.
The inclusion of these contributions is achieved by solving
again Eqs.~(\ref{o1}), (\ref{o2}) and (\ref{o3}) but with Eq.~(\ref{newR}).
There are of course other choices for $\Gamma_p$ than the present one,
which will be discussed in Sec.~\ref{discussion}. 
We find that the present scheme
is relatively easy to implement numerically and to generalize to the next order.
To get an improvement of the present approximation, we just need to calculate 
the contribution from the next order skeleton diagram shown in Fig.~\ref{npn}
and include it into Eq.~(\ref{newR}) in a similar way.

\section{Results}
\label{results}

\begin{figure}
\includegraphics[width=0.45\textwidth]{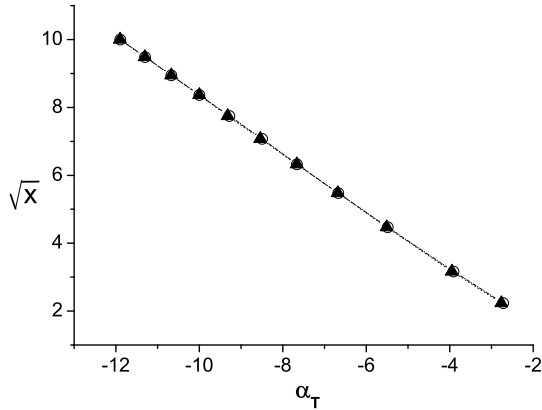}%
\caption{The renormalized propagator $\sqrt{x}\sim 1/\alpha_R$ 
as a function of the temperature parameter $\alpha_T$. The filled triangles 
and the circles are obtained with and without 
the non-parquet diagrams, respectively. The dashed line is a guide
for the eye.}
\label{alpha_t}
\end{figure}

\begin{figure}
\includegraphics[width=0.45\textwidth]{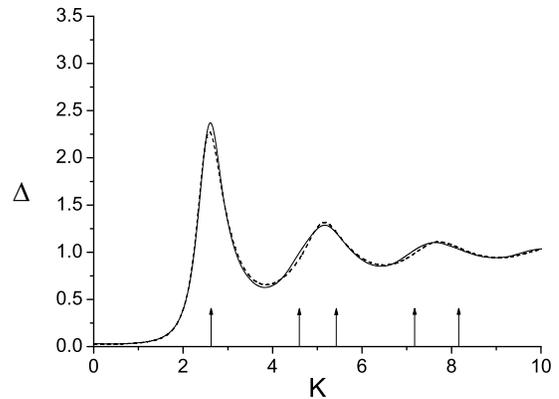}%
\caption{The structure factor $\Delta$ of the two-dimensional vortex liquid 
as a function of the dimensionless momentum $K=k/\mu$ at
$\alpha_T\simeq -7.66$. The solid line is from the calculation including 
the non-parquet diagrams, while the dashed lines is that of the parquet
approximation. The arrows indicate the positions of the RLV of the triangular lattice.}
\label{x40}
\end{figure}

\begin{figure}
\includegraphics[width=0.45\textwidth]{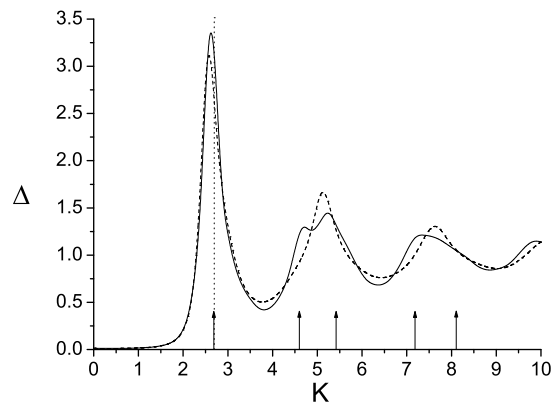}%
\caption{The structure factor at $\alpha_T\simeq -11.9$. The solid and dashed lines
are the same as in Fig.~\ref{x40}.
The expected position of the
first peak from the RLV is presented as a dotted line for a guide for the eye.}
\label{x100}
\end{figure}

As explained in the previous section,
we solve numerically the self-consistent equations 
(\ref{o1}), (\ref{o2}) and (\ref{o3}) together with (\ref{newR})
for $\Gamma({\bf k})$. The vertex function $\Gamma_p({\bf k})$
used in (\ref{newR}) is obtained by solving once
Eqs.~(\ref{o1}), (\ref{o2}) and (\ref{o3}) with $R({\bf k})=\Gamma_B({\bf k})$.
All these numerical procedures are performed for given value of $x$.
The temperature parameter $\alpha_T$ is determined from Eq.~(\ref{dyson}).
Since we are considering the vortex liquid phase, we only consider
a rotationally symmetric case where all the functions depend on
a dimensionless momentum $K$ with ${\bf K}={\bf k}/\mu$.
The integral equations are solved numerically by iteration starting 
from an appropriate choice of initial $\Gamma({\bf k})$. 
The convergence of the iteration
can be improved when the solution at slightly smaller value of $x$ is used
as the initial choice. We obtain the solutions for given values of $x$
up to $x=100$, which is the largest value we considered.
It corresponds approximately to $\alpha_T=-11.9$ (see Fig.~\ref{alpha_t}). 
As we go down to lower temperatures, we have to 
increase the $k$-space cutoff to accommodate the peaks appearing 
at large $k$ in the structure factor and 
decrease the grid size at the same time to capture the sharp first peak 
(see Figs.~\ref{x40} and \ref{x100}). The number of iteration
needed to get a convergence increases as the temperature is lowered.
At the lowest temperature we considered, we needed about 1000 iterations
to get a convergence.
All these factors limit the temperature range where the numerical 
solution can be obtained.

Figure \ref{alpha_t} shows the relationship between the renormalized
propagator $\sqrt{x}\sim\alpha_R^{-1}$
and the temperature parameter $\alpha_T$. The values are
compared with those obtained when
only the parquet diagrams are considered. We can see that
there are essentially no difference in the two cases. 
Note that in the present analysis,
we have extended the previous parquet approximation
results \cite{ym} to lower temperatures $\alpha_T\simeq -11.9$.
The dimensionless renormalized propagator $\sqrt{x}$ is directly proportional 
to the thermodynamic quantities like the magnetization and
the entropy of the vortex liquid system.
Another interesting thermodynamic quantity 
is the generalized Abrikosov ratio defined by
$\beta_A(x)\equiv \overline{\langle|\Psi({\bf r})|^4\rangle} /
[\overline{\langle|\Psi({\bf r})|^2\rangle}]^2 $, where the bar denotes the spatial 
average. As shown in Ref.~\onlinecite{ym}, this quantity decreases as the temperature
is lowered, and approaches 1.16, the value for a triangular lattice as $x\to \infty$
or $\alpha_T\to -\infty$. It is related to $\Gamma({\bf k})$ in such a way that
the relation (\ref{dyson}) can be rewritten as $\alpha_T =(1-x\beta_A(x))/\sqrt{x}$.
Therefore, the generalized Abrikosov ratio even in the presence of the non-parquet 
contributions shows the similar behavior to the result obtained in the parquet
approximation \cite{ym}. From these results, we may conclude that
non-parquet contributions make little difference on the
thermodynamic quantities. 
As we will see below, however, effects of non-parquet diagrams appear in
the crystalline order developing in the vortex liquid.

We calculate the structure factors (\ref{dfv}) for the two-dimensional vortex
liquid at various temperatures as shown
in Figs.~\ref{x40} and \ref{x100}. They are compared with those
obtained in the parquet approximation. 
The structure factors for both cases look almost the same down to low temperatures 
$\alpha_T\simeq -7$. As the temperature is lowered further, however,
we find that the 
first peak becomes slightly larger and sharper when the non-parquet
diagrams are included (see Fig.~\ref{x40}). This trend continues further down to
lower temperatures. When the temperature is lowered below $\alpha_T\simeq -10$,
the second peak begins to split into two peaks (see Fig.~\ref{x100}).
This can be interpreted as the non-parquet contributions capturing the 
growing crystalline order in the vortex liquids more effectively.
The peaks developing in the structure factor correspond to the positions 
of the reciprocal lattice vectors (RLV) of the triangular lattice. 
The RLV ${\bf G}$ can be represented in a dimensionless form as
${\bf G}/\mu=G_0\;(m\eta,n-m\zeta)$ using a set of integers 
$m$ and $n$, where, for the triangular lattice, $\eta=\sqrt{3}/2$,
$\zeta=1/2$ and $G_0=\sqrt{2\pi/\eta}\simeq 2.694$. 
Therefore the lengths of the RLV can be grouped into
$|{\bf G}|/\mu=c_i G_0$, $i=1,2,3\ldots$, where 
$c_1=1, c_2=\sqrt{3}, c_3=2, c_4=\sqrt{7}, c_5=3$, etc.
The first peak in the structure factor is located near $K=G_0$.
Comparing the results in Fig.~\ref{x100} for the cases
with and without the non-parquet diagrams, we can see that when the non-parquet
contributions are included, the first peak is closer to its expected position.   
Since, for a triangular lattice, 
the second and third sets of RLV are relatively closely spaced, they 
appear as one peak within the parquet approximation. Figure \ref{x100}
shows that the resolved peaks are located near the expected
positions of the RLV when the non-parquet contributions are included. 
The situation is similar for the closely-spaced fourth and fifth peaks.
Although the peaks are not resolved at the minimum temperature we have studied,
we can clearly see a tendency compared to the parquet results.

\section{Discussion and Conclusion}
\label{discussion}

In summary we have generalized the 
parquet approximation for the two-dimensional vortex liquid systems
by demonstrating that the non-parquet contributions
can be included systematically into the nonperturbative calculation
of the correlation functions.
The crystalline order developing in the vortex liquid as the temperature
is lowered is captured more effectively with the inclusion 
of the non-parquet diagrams in the sense that the peaks in the structure
factor corresponding to the RLV of the triangular lattice become resolved. 

In Ref.~\onlinecite{ym}, within the parquet approximation,
no finite temperature phase
transition has been observed in the two-dimensional vortex
liquid as the temperature is lowered. The length scale 
characterizing the growing crystalline order in the vortex
liquid was determined from the width of the first peak in 
the structure factor. It was shown that
this length scale grows like $\sqrt{x}\sim |\alpha_T|$
in the low temperature limit. Since the first peaks in 
the structure factor change only slightly with the inclusion 
of the non-parquet diagrams, these conclusions drawn 
from the parquet approximation remain valid. The detailed 
structure factor has a more accurate form in the present
generalization.

As mentioned in Sec.~\ref{diagrams}, there are other possible
generalizations of the parquet approximation than the present one.
They correspond to different choices for the vertex $\Gamma_p({\bf k})$
used in Eq.~\ref{J}. We can, for example, take 
$\Gamma_p$ equal to the full vertex $\Gamma({\bf k})$ not just the sum of
all the parquet diagrams. 
Obviously the diagrams 
in Fig.~\ref{np} in this scheme contain more diagrams than in the
present one. We have attempted to solve numerically
the set of equations (\ref{o1}), (\ref{o2}), (\ref{o3})
and (\ref{newR}) with (\ref{J}) when $\Gamma_p=\Gamma$.  
At relatively high temperatures, we find there is very little
difference in correlations of the vortex liquid between this
scheme and the present one. But the numerical iteration in this
case involves repeated evaluation of the integral in Eq.~\ref{J}
at every step of the iteration, which considerably slows down 
the whole calculation. We find that it is not practical to use
this approximation below $\alpha_T\simeq -3$. We believe that
the effect of the non-parquet diagrams is already captured in
the present approximation scheme.



\begin{acknowledgments}
This work was supported by Konkuk University in 2005.
\end{acknowledgments}


\end{document}